\documentclass{article}

\usepackage[utf8]{inputenc} 
\usepackage[T1]{fontenc}    
\usepackage{hyperref}       
\usepackage{url}            
\usepackage{booktabs}       
\usepackage{amsfonts}       
\usepackage{nicefrac}       
\usepackage{microtype}      
\usepackage{lipsum}
\usepackage{graphicx}

\usepackage{moreverb,url}
\usepackage{amsmath}
\usepackage[american]{babel}
\usepackage{tipa}
\usepackage{pstricks}
\usepackage{bm}
\usepackage{fancyhdr, indentfirst}
\usepackage{lscape}
\usepackage{amssymb}
\usepackage{multirow}
\usepackage{geometry}
\usepackage{lmodern}
\usepackage{float}
\usepackage{amsthm}
\usepackage{setspace}
\usepackage{amssymb}
\usepackage{threeparttable}
\usepackage{amsthm}

\newtheorem{thm}{Theorem}

\newcommand\independent{\protect\mathpalette{\protect\independenT}{\perp}}
\def\independenT#1#2{\mathrel{\rlap{$#1#2$}\mkern2mu{#1#2}}}

\usepackage{placeins}
\usepackage[ruled]{algorithm2e}
\usepackage{booktabs}
\usepackage[marginal]{footmisc}

\title{Empirical Likelihood Weighted Estimation of Average Treatment Effects in Randomized Clinical Trials}

\begin{document}
\begin{center}{
\Large \bf     Empirical Likelihood Weighted Estimation of  Average Treatment Effects}
         \\
                \vspace{2mm}
        {\large Yuanyao Tan$^1$, Xialing Wen$^{1}$, Wei Liang$^1$ and Ying Yan$^{1*}$}

        {\small $^1$ School of Mathematics, Sun Yat-sen University, Guangzhou, China\\
        $^*$ The Corresponding Author. Email: \textit{yanying7@mail.sysu.edu.cn}}
         \end{center}


\begin{abstract}
There has been growing attention on how to effectively and objectively
use covariate information when the primary goal is to estimate the
average treatment effect (ATE) in randomized clinical trials (RCTs).
In this paper, we propose an effective weighting approach to extract
covariate information based on the empirical likelihood (EL) method.
The resulting two-sample empirical likelihood weighted (ELW) estimator
includes two classes of weights, which are obtained from a constrained
empirical likelihood estimation procedure, where the covariate information
is effectively incorporated into the form of general estimating equations.
Furthermore, this ELW approach separates the estimation of ATE from
the analysis of the covariate-outcome relationship, which implies
that our approach maintains objectivity. In theory, we show that the
proposed ELW estimator is semiparametric efficient. We extend our
estimator to tackle the scenarios where the outcomes are missing at
random (MAR), and prove the double robustness and multiple robustness
properties of our estimator. Furthermore, we derive the semiparametric
efficiency bound of all regular and asymptotically linear semiparametric ATE estimators under MAR mechanism and prove that our proposed estimator attains this bound. We conduct simulations
to make comparisons with other existing estimators, which confirm
the efficiency and multiple robustness property of our proposed ELW estimator. An application to
the AIDS Clinical Trials Group Protocol 175 (ACTG 175) data is conducted.
\end{abstract}


\section{Introduction}
The RCTs aim to compare various treatments
when the subjects are randomized to enter different treatment groups.
The ATE is commonly used in RCTs as it measures the difference in the mean outcomes between two treatment
groups. A natural estimator of ATE is the difference in the empirical
average outcomes between the treatment group and the control group;
it is unbiased due to randomization. When there exists possible
association between the primary outcome and the extensively collected
baseline covariates in RCTs, the precision of the ATE estimator may
be improved by adjusting for the effect of covariates.
There exists a voluminous literature dealing with covariate adjustment \cite{senn1989covariate,koch1998issues,lesaffre2003note,leon2003semiparametric,davidian2005semiparametric,tsiatis2008covariate}
to improve the precision of the estimator and increase statistical
power. However, it also contains considerable debate regarding the
appropriateness of covariate adjustment \cite{lesaffre2002variability,pocock2002subgroup}.
Concerns mainly focus on the potential bias in treatment effect estimation,
which is caused by post hoc selection of covariates and by allowing
investigators to go on a ``fish expedition'' to find models with
the most significant estimate of treatment effect. To address such
concerns, a certain number of approaches are proposed to maintain
objectivity when adjusting covariates in randomized trials. By utilizing
the semiparametric theory, Tsiatis et al.  \cite{tsiatis2008covariate}
proposed a systematic method to objectively incorporate covariate
effects while exploiting the relationship between covariates and response
outcomes, by positing two separate working regression models for the
data from the two treatment groups, leading to an increase in precision.
Besides, Shen et al.\cite{shen2014inverse} and Williamson
et al.\cite{williamson2014variance} put forward two two-stage
estimation procedures for covariate adjustment based on the inverse
probability weighting (IPW) method. They tried to adjust for covariates
by estimating the propensity score without using outcome data to ensure
objectivity.

The empirical likelihood (EL) method is also an appealing method to
adjust for baseline covariates in the estimation of ATE\cite{zhang2018empirical,huang2008empirical}.
Since Owen \cite{owen1988empirical} first proposed the EL method
as a nonparametric likelihood procedure to construct confidence intervals
for the mean and other parameters, there have been numerous advances
bringing the application of EL to many research areas. We refer interested
readers to Owen's 2001 monograph\cite{owen2001empirical} for further details. An important
work done by Qin and Lawless \cite{qin1994empirical} showed
that the EL method can effectively incorporate side information in
the form of general estimating equations (GEE) into inference through
constrained maximization of the empirical likelihood function. Their
work inspired some researchers to utilize EL to make covariate adjustments in RCTs and related clinical designs.

Zhang \cite{zhang2018empirical} considered two unbiased estimating
functions that automatically decouple the estimation of ATE from the regression
modeling of covariate-outcome relationship and their resulting estimator
can reach the same efficiency as the existing efficient adjusted estimators
do \cite{tsiatis2008covariate}. Considering the estimation of ATE in pretest-posttest studies, Huang et al.
\cite{huang2008empirical} proposed an empirical likelihood-based
estimation procedure that can incorporate the common baseline covariate
information to improve efficiency.

When the outcome is missing in some of the observations in RCTs, great
uncertainty and possible bias in the estimation of ATE may exist.
Here, we mainly focus on situations with data missing at random (MAR),
i.e., conditioning on the covariates and responses, the missing outcomes
depend only on the covariates \cite{little2002statistical}. In order
to correct for the bias caused by missingness, various methods have
been proposed, including the weighting methods originated by Horvitz
and Thompson \cite{horvitz1952generalization}.
In the context of the pretest-posttest study with missing data, Davidian
et al. \cite{davidian2005semiparametric} studied a class of
consistent semiparametric estimators for the treatment effect and
identified the most efficient one based on the semiparametric theory.
However, the construction of the semiparametric efficient estimator
depends on whether the underlying relationship between the outcome
and covariates is correctly specified. This estimator can be much
less efficient if the ``working regression model'' and the true
regression model are not close to each other, especially when the
dimension of covariates is high.

Recently, empirical likelihood methods have been received growing
attention to missing data problems for its attractive data-driven
feature and nice robustness property. Qin and Zhang \cite{qin2007empirical}
proposed an empirical likelihood-based approach to estimate the mean
response under the MAR assumption, the resulting estimators enjoy
the double-robustness property, i.e., the estimator of the mean response
is asymptotically unbiased if either the underlying propensity score
or the underlying regression function is correctly specified. Huang
et al. \cite{huang2008empirical} applied the EL method to estimate
the treatment effect in the pretest-posttest setting with missing
data; they considered counterfactual missing data to estimate EL weights
which were not considered by Qin and Zhang \cite{qin2007empirical}.
Chen et al. \cite{chen2015imputation} proposed an imputation-based
empirical likelihood approach to adjust for baseline information and
dealt with the responses in pretest-posttest studies which are
missing by design. However, none of their work defines the estimator of ATE as the difference
of two weighted outcomes with two separate classes of weights obtained
from constrained maximization of the empirical likelihood function.

In this article, we propose a new approach to incorporate covariate
information into the estimation of ATE using the EL method. Inspired
by the work of Wu and Yan \cite{wu2012empirical}, we construct
our estimator by separately weighting the outcomes of two samples,
where the weights are estimated to carry covariate information through
moment constraints which implicitly utilize randomization inherited
in RCTs. These constraints focus solely on covariates and treatment
assignments but not on the outcomes. To exploit the relationship between
the covariates and the outcomes, we posit two models for each treatment
group through parametric regression or identity function, then use them
in the moment constraints. Therefore, we separate the modeling of
the covariate-outcome relationship from the ATE estimation, making
the covariate adjustment procedure objective. Also, we extend our
approach to the scenarios where the outcomes are partly missing. In this case, we prove the double robustness, multiple robustness and semiparametric efficiency for our proposed estimator.

Zhang's \cite{zhang2018empirical} recent work focused on estimating the ATE by adding the parameter of interest and the covariate information in the estimating functions and deriving the asymptotic form of the ATE estimator using the empirical likelihood theory. In contrast, we first construct the two-sample ELW estimator for ATE with the estimated weights, which are designed to carry the covariate information based on the EL method; then we discuss the asymptotic property for the proposed estimators. Furthermore, Zhang's method didn't consider the possible missingness of the outcome data and the corresponding robustness properties in this case, which we take into account in this paper.

When dealing with missing outcomes, we follow the work of Qin and
Zhang \cite{qin2007empirical} by adding two moment constraints
to take missing mechanism into account. However, we propose to use
the combined information from the treatment group and the control
group to construct the two moment constraints for the propensity scores,
whereas Qin and Zhang treated the two constraints separately. Intuitively,
our estimator is more efficient. In fact, we prove that our estimator
is semiparametric efficient. Furthermore, we prove that our estimator is doubly robust and multiply robust \cite{han2013estimation,han2014multiply}.

In Section \ref{sec:prop}, we introduce the proposed weighted empirical likelihood
estimator and show the extensions of our method to incorporate missing
outcomes and enhance multiple robustness. We show the details of the
practical implementation of the proposed method in Section \ref{sec:opt}. In section
\ref{sec:simu}, the performance of our method is evaluated by a series of simulations
and an application to ACTG175 data. We draw conclusions in Section \ref{sec:concl}. Proofs are presented in the supplementary material.

\section{Proposed Methodology}
\label{sec:prop}
In Section \ref{sec:nomis}, we describe our method in the standard RCTs where there is no missingness in the outcomes.
In Section \ref{sec:mis},
we consider the scenario where outcomes are partly missing  under
the missing at random mechanism. Furthermore, we apply multiple working models
to enhance robustness in the estimation, which leads to the multiple
robustness property described in Section \ref{sec:Multiple Robustness}. 

\subsection{RCTs without missing outcomes}
\label{sec:nomis}
Consider a two-arm randomized clinical trial comparing the treatment
group and the control group. Let $W$ be a binary variable with $W=1$
if treated and $W=0$ if controlled. Define $\delta=P(W=1)$ to be
the probability of being treated and assume  $0<\delta<1$. Let $Y_{0}$ ($Y_{1}$) be the outcome
of a subject from the control (treatment) group. We define the outcome
for each subject in a unified way as $Y=WY_{1}+(1-W)Y_{0}$. Denote
$X_{l\times1}$ to be a $l$-dimensional vector of baseline covariates.
Under randomization in the RCTs, treatment assignment and baseline
covariates are independent, i.e., $W\independent X$. Therefore, $X|W=1$
and $X|W=0$ have the same distribution as that of the covariate $X$
in the entire sample, i.e.,  $\rho_{1}(x)=\rho_{0}(x)=\rho(x)$ where
we define $\rho_{1}(x)$ and $\rho_{0}(x)$ as the probability density
function of the covariate $X$ in the treatment group and the control
group, respectively, and $\rho(x)$ as that of the covariate
$X$ in the entire sample. The observed data of the treatment group
$\{(X_{1i},Y_{1i}),i=1,\cdots,m\}$ are independent and identically
distributed (i.i.d.). Likewise, the observed data of the control group
$\{(X_{0j},Y_{0j}),j=1,\cdots,n\}$ are i.i.d.. Let $N=m+n$ be the
total size of the two samples. Denote $\mu_{1}=E(Y_{1})$ and $\mu_{0}=E(Y_{0})$.
We are interested in estimating ATE, given by $\theta=\mu_{1}-\mu_{0}=E(Y_{1})-E(Y_{0})$,
from the observed data.

We introduce an empirical likelihood method to effectively incorporate
covariate information when estimating ATE. Let $f_{1}(x,y_{1})$ be
the joint density function of $(X,Y_{1})$ and $f_{0}(x,y_{0})$ be
the joint density function of $(X,Y_{0})$. Let $p_{i}=f_{1}(X_{1i},Y_{1i})$
for $i=1,\cdots,m,$ and $q_{j}=f_{0}(X_{0j},Y_{0j})$ for $j=1,\cdots,n,$
be the probability mass at point $(X_{1i},Y_{1i})$ and $(X_{0j},Y_{0j})$,
respectively. The nonparametric likelihood for the observed data is
\begin{equation}
\prod_{i=1}^{m}p_{i}\prod_{j=1}^{n}q_{j}.\label{nonlik}
\end{equation}

We propose to obtain the estimators of the $p_{i}$'s and $q_{j}$'s,
by maximizing the likelihood $(\ref{nonlik})$ subject to the following
constraints
\begin{eqnarray}
\sum_{i=1}^{m}p_{i} & = & 1,\ \ p_{i}\geq0,\ \ i=1,\cdots,m, \label{eq:nomis1}\\
\sum_{j=1}^{n}q_{j} & = & 1,\ \ q_{j}\geq0,\ \ j=1,\cdots,n, \label{eq:nomis2}\\
\sum_{i=1}^{m}p_{i}g(X_{1i}) & = & \bar{g}
, \label{eq:nomis3}\\
\sum_{j=1}^{n}q_{j}h(X_{0j}) & = & \bar{h}
,\label{eq:nomis4}
\end{eqnarray}
where $\bar{g}=\frac{1}{N}\left\{ \sum_{i=1}^{m}g(X_{1i})+\sum_{j=1}^{n}g(X_{0j})\right\}$ and $h(x)=\frac{1}{N}\left\{ \sum_{i=1}^{m}h(X_{1i})+\sum_{j=1}^{n}h(X_{0j})\right\}$.  The $g(x)$ and $h(x)$ are arbitrary $r_1$-dimensional and $r_0$-dimensional functions, respectively. We take $r_1\geq1$ and $r_0\geq1$ as two integers. The constraints $(\ref{eq:nomis1})$ and $(\ref{eq:nomis2})$ ensure
that $p_{i}$'s and $q_{j}$'s are the empirical probabilities. The
latter two constraints (\ref{eq:nomis3}) and (\ref{eq:nomis4}) are the empirical versions of two equations $E\{g(X)|W=1\}=E\{g(X)\}$
and $E\{h(X)|W=0\}=E\{h(X)\}$, which utilize the fact that the two groups
have identical baseline covariate distributions due to the randomization
procedure in the RCTs. Since the constraints for the $p_{i}$'s do
not involve any of the $q_{j}$'s and vice versa, we can estimate
the $p_{i}$'s and the $q_{j}$'s separately as two optimization problems.
Note that $g(x)$ and $h(x)$ are known functions, for instance, they can
be identity functions, linear functions of the covariates, etc.

Since the above optimization problem is a strictly convex problem, there exists
an unique global maximum under some mild conditions, including the
convex hull condition that $\bar{g}$ and $\bar{h}$ are inside the
convex hull of $\{g(X_{1i}),i=1,\cdots,m\}$ and $\{h(X_{0j}),j=1,\cdots,n\}$,
respectively \cite{owen2001empirical}. The solutions can be obtained
by using the method of Lagrange multipliers (details are shown in
Section \ref{sec:opt}):
\begin{align*}
\hat{p}_{i}&=\frac{1}{m}\frac{1}{1+\lambda_{1}^{\top}\{g(X_{1i})-\bar{g}\}},i=1,\cdots,m,\\
\hat{q}_{j}&=\frac{1}{n}\frac{1}{1+\lambda_{2}^{\top}\{h(X_{0j})-\bar{h}\}},j=1,\cdots,n,
\end{align*}
where $\lambda_{1}$ and $\lambda_{2}$ are the Lagrange multipliers determined by
\[
\frac{1}{m}\sum_{i=1}^{m}\frac{g(X_{1i})-\bar{g}}{1+\lambda_{1}^{\top}\{g(X_{1i})-\bar{g}\}}=0,\quad\frac{1}{n}\sum_{j=1}^{n}\frac{h(X_{0j})-\bar{h}}{1+\lambda_{2}^{\top}\{h(X_{0j})-\bar{h}\}}=0,
\]
respectively. Our proposed two-sample empirical likelihood weighted
(ELW) estimator is
\[
\hat{\theta}=\sum_{i=1}^{m}\hat{p}_{i}Y_{1i}-\sum_{j=1}^{n}\hat{q}_{j}Y_{0j},
\]
which is consistent for the ATE under suitable regularity conditions due to the following theorem.
\begin{thm}
	\label{consisthm} As $N\xrightarrow{}\infty$, $m/N\xrightarrow{}\delta>0$ and $n/N\xrightarrow{}1-\delta>0$, $\hat{\theta}$ is a consistent estimator for $\theta$.
\end{thm}
\noindent
The regularity conditions and the proofs of Theorem \ref{consisthm} and other theorems in the article are provided in the supplementary material.

Usually, we take $g(x)$ and $h(x)$ as two parametric outcome regression models  $\widetilde{g}(x;\beta_{1})$
and $\widetilde{h}(x;\beta_{0})$ to approximate $\widetilde{g}(x)=E(Y|W=1,X=x)$ and $\widetilde{h}(x)=E(Y|W=0,X=x)$. Note that taking $g(x)$ and $h(x)$ as the identity functions can be seen as adding multiple moment constraints for multiple parametric outcome regression models, each of which only involves one covariate.
In practice, we estimate $\beta_{1}$ and $\beta_{0}$ by their corresponding estimators $\hat{\beta}_{1}$ and $\hat{\beta}_{0}$, which are obtained by fitting two parametric outcome regression models  $\widetilde{g}(x;\beta_{1})$
and $\widetilde{h}(x;\beta_{0})$ separately using the least square method. According to White
\cite{white1982maximum}, under suitable regularity conditions, $\hat{\beta}_{1}\xrightarrow{}{}\beta_{1*}$ and $\hat{\beta}_{0}\xrightarrow{}{}\beta_{0*}$ in probability as $N\rightarrow\infty$ where $*$ denote
the corresponding values of the parameters that minimizes the Kullback-Leibler distance
from the probability distribution function based on the postulated model to the true one that generates the data. Generally, $\widetilde{g}(x;\beta_{1*})\neq E(Y|W=1,X=x)$ unless $\widetilde{g}(x;\beta_{1})$ is correctly specified and $\widetilde{h}(x;\beta_{0*})\neq E(Y|W=0,X=x)$ unless $\widetilde{h}(x;\beta_{0})$ is correctly specified. In addition, we have $\bar{g}(\hat{\beta}_{1})\xrightarrow{}E\{\widetilde{g}(\beta_{1*})\}$ and $\bar{h}(\hat{\beta}_{0})\xrightarrow{}E\{\widetilde{h}(\beta_{0*})\}$  in probability as $N\rightarrow\infty$. Here, we set $g(x)$ and $h(x)$ in (\ref{eq:nomis3}) and (\ref{eq:nomis4}) to be $g(x)=\widetilde{g}(x;\hat{\beta}_1)$ and $h(x)=\widetilde{h}(x;\hat{\beta}_0)$. The following theorem gives the asymptotic distribution for $\hat{\theta}$ in this case.

\begin{thm} \label{nomisthm} As $N \rightarrow \infty$,  $N^{1/2}(\hat{\theta}-\theta)$ follows	an asymptotically normal distribution with mean 0 and variance $\text{var}\{\varphi(Y,X,W)\}$
	with the influence function
	\[
	\begin{aligned}\varphi(Y,X,W)= & \frac{W}{\delta}(Y-\mu_{1})-\frac{W-\delta}{\delta}C_{1}^{\top}D_{1}^{-1}\left[\widetilde{g}(X;\beta_{1*})-E\{\widetilde{g}(X;\beta_{1*})\}\right]\\
	& -\frac{1-W}{1-\delta}(Y-\mu_{0})+\frac{W-\delta}{1-\delta}C_{0}^{\top}D_{0}^{-1}\left[\widetilde{h}(X;\beta_{0*}) 
	-E\{\widetilde{h}(X;\beta_{0*})\}\right],
	\end{aligned}
	\]
	where
	\[
	\begin{aligned}C_{1} & =E\left(\frac{W}{\delta}(Y-\mu_{1})\left[\widetilde{g}(X;\beta_{1*})-E\{\widetilde{g}(X;\beta_{1*})\}\right]\right),\\
	C_{0} & =E\left(\frac{1-W}{1-\delta}(Y-\mu_{0})\left[\widetilde{h}(X;\beta_{0*})-E\{\widetilde{h}(X;\beta_{0*})\}\right]\right),\\
	D_{0} & =E\left(\left[\widetilde{h}(X;\beta_{0*})-E\{\widetilde{h}(X;\beta_{0*})\}\right]^{\otimes2}\right),\\
	D_{1} & =E\left(\left[\widetilde{g}(X;\beta_{1*})-E\{\widetilde{g}(X;\beta_{1*})\}\right]^{\otimes2}\right).
	\end{aligned}
	\]
\end{thm}

When $\widetilde{g}(x;\beta_{1})$ and $\widetilde{h}(x;\beta_{0})$ are correctly specified; namely, $\widetilde{g}(x;\beta_{1*})=\widetilde{g}(x)=E(Y|W=1,X=x)$ and $\widetilde{h}(x;\beta_{0*})=\widetilde{h}(x)=E(Y|W=0,X=x)$,
we have
\[
\begin{aligned}\varphi_{opt}(Y,X,W)= & \frac{W}{\delta}(Y-\mu_{1})-\frac{W-\delta}{\delta}\left\{ E(Y|X,W=1)-\mu_{1}\right\} \\
& -\frac{1-W}{1-\delta}(Y-\mu_{0})-\frac{W-\delta}{1-\delta}\left\{ E(Y|X,W=0)-\mu_{0}\right\},
\end{aligned}
\]
which is the efficient influence function for regular and asymptotically linear (RAL) estimators of $\theta$ in RCTs described by Tsiatis et. al
\cite{tsiatis2007semiparametric}. In this case, $\text{var}\{\varphi_{opt}(Y,X,W)\}$
is the asymptotic variance of $N^{1/2}(\hat{\theta}-\theta)$, which
equals to the semiparametric efficiency bound. This observation leads to the following
theorem on the efficiency of $\hat{\theta}$. \begin{thm} \label{nomisbound}
	When $\widetilde{g}(x;\beta_{1})$ is correctly specified for $E(Y|W=1,X=x)$
	and $\widetilde{h}(x;\beta_{0})$ is correctly specified for $E(Y|W=0,X=x)$, the asymptotic variance
	of $\hat{\theta}$ attains the semiparametric efficiency bound. \end{thm}

According to Theorem \ref{nomisthm}, $\hat{\theta}$ is still consistent even if $\widetilde{g}(x;\beta_{1})$ and $\widetilde{h}(x;\beta_{0})$ are not correctly
specified, but it is not semiparametric efficient due to Theorem \ref{nomisbound}. Details for the proofs are shown in the supplementary material.

\subsection{RCTs with missing outcomes}
\label{sec:mis}

In this section, we follow the work of Qin and Zhang  \cite{qin2007empirical}
to take missing outcomes into account. However, their work tackled
the one-sample case; we extend their work to the RCT data and take
randomization into account when we construct our empirical likelihood
estimator. Suppose $Y$ is missing for some subjects, and the baseline
covariates $X$ are always observed. Let $R_{1}$ ($R_{0}$) be the
missing indicator for treatment group (control group) that takes value
0 if $Y_{1}$ ($Y_{0}$) is missing and 1 otherwise. The observed
data are $\{(R_{1i}Y_{1i},R_{1i},X_{1i}),i=1,\cdots,m;(R_{0j}Y_{0j},R_{0j},X_{0j}),j=1,\cdots,n\}$.
We reformulate the data into a two-sample setting as
\begin{center}
	$\{(Y_{1i},X_{1i});i=1,\cdots,m_{0}\}$, $Y_{1i}$ is observed in
	the treatment group;\\
	$\{(?,X_{1i});i=m_{0}+1,\cdots,m\}$, $Y_{1i}$ is missing in the
	treatment group;\\
	$\{(Y_{0j},X_{0j});j=1,\cdots,n_{0}\}$, $Y_{0j}$ is observed in
	the control group;\\
	$\{(?,X_{0j});j=n_{0}+1,\cdots,n\}$, $Y_{0j}$ is missing in the
	control group.\\
	\par\end{center}

For unified notation, we define $R=WR_{1}+(1-W)R_{0}$, $Y=WRY_{1}+(1-W)RY_{0}$.
Then the observed data can be written as $\{(Y_{k},X_{k},R_{k},W_{k}),k=1,\cdots,N\}$.
We impose the common MAR mechanism\cite{little2002statistical};
that is, $P(R=1|Y,X,W)=P(R=1|X,W)$. Denote the missing probabilities for the treatment and control groups
as $\pi_{1}(x)=P(R=1|W=1,X=x)$ and $\pi_{0}(x)=P(R=1|W=0,X=x)$, respectively. We specify $\pi_{1}(x;\alpha_{1})$
as a parametric model to approximate $\pi_{1}(x)$, likewise, $\pi_{0}(x;\alpha_{0})$
is a parametric model to approximate $\pi_{0}(x)$. The $\alpha_{1}$
and $\alpha_{0}$ are given as the unknown vector parameters. In practise, we usually model model the propensity scores $\pi_{w}(x)$, $w=0,1$, with logistic regression models.

Our interest is still to estimate $\theta=E(Y_{1})-E(Y_{0})$ in the
presence of missingness in the outcomes. Here our proposed estimator
is $\hat{\theta}_{\text{mis}}=\sum_{i=1}^{m_{0}}{\hat{p}_{i}Y_{1i}}-\sum_{j=1}^{n_{0}}{\hat{q}_{j}Y_{0j}}$
where $\hat{p}_{i}$'s and $\hat{q}_{j}$'s are obtained by maximizing
the following nonparametric likelihood
\begin{equation}
\label{eq:nonlik2}
    \prod_{i=1}^{m_{0}}p_{i}\prod_{j=1}^{n_{0}}q_{j}
\end{equation}

subject to
\begin{eqnarray}
\sum_{i=1}^{m_{0}}{p_{i}} & = & 1,\ \ p_{i}\geq0,\ \ i=1,\cdots,m_0, \nonumber \\
\sum_{j=1}^{n_{0}}{q_{j}} & = & 1,\ \ q_{j}\geq0,\ \ j=1,\cdots,n_0, \nonumber \\
\sum_{i=1}^{m_{0}}p_{i}\pi_{1}(X_{1i};\hat{\alpha}_{1}) & = & \overline{\pi}_{1},\label{eq:mis1}\\
\sum_{j=1}^{n_{0}}q_{j}\pi_{0}(X_{0j};\hat{\alpha}_{0}) & = & \overline{\pi}_{0},\label{eq:mis2}\\
\sum_{i=1}^{m_{0}}p_{i}g(X_{1i}) & = & \bar{g},\label{eq:mis3}\\
\sum_{j=1}^{n_{0}}q_{j}h(X_{0j}) & = & \bar{h},\label{eq:mis4}
\end{eqnarray}
where $\overline{\pi}_{1}=\frac{1}{N}\left\{ \sum_{i=1}^{m}\pi_{1}(X_{1i};\hat{\alpha}_{1})+\sum_{j=1}^{n}\pi_{1}(X_{0j};\hat{\alpha}_{1})\right\} $,
$\overline{\pi}_{0}=\frac{1}{N}\left\{ \sum_{i=1}^{m}\pi_{0}(X_{1i};\hat{\alpha}_{0})+\right.\left.\sum_{j=1}^{n}\pi_{0}(X_{0j};\hat{\alpha}_{0})\right\} $.
The first two constraints guarantee that $p_{i}$'s and $q_{j}$'s
are empirical probabilities. The constraints  (\ref{eq:mis1}) and (\ref{eq:mis2}) reflect
the selection bias according to Qin and Zhang  \cite{qin2007empirical}. Similarly,
the latter two constraints (\ref{eq:mis3}) and (\ref{eq:mis4}) utilize covariate information through functions $g(x)$ and $h(x)$. As described in Section \ref{sec:nomis}, we set $g(x)=\widetilde{g}(x;\hat{\beta}_1)$ and $h(x)=\widetilde{h}(x;\hat{\beta}_0)$, which are the parametric estimations for
$\widetilde{g}(x)=E(Y|W=1,X=x)$ and $\widetilde{h}(x)=E(Y|W=0,X=x)$, respectively. Here, we have the
following result on the consistency of $\hat{\theta}_{\text{mis}}$.

\noindent \begin{thm} \label{doublerobustthm}
	$\hat{\theta}_{\text{mis}}$ is
	consistent for $\theta$ as $N \xrightarrow{} \infty$ if both the following conditions are satisfied:
	i) either $\pi_{1}(x;\alpha_{1})$ is correctly specified for $\pi_{1}(x)$
	or $\widetilde{g}(x;\beta_{1})$ is correctly specified for $E(Y|W=1,X=x)$; ii)
	either $\pi_{0}(x;\alpha_{0})$ is correctly specified for $\pi_{0}(x)$
	or $\widetilde{h}(x;\beta_{0})$ is correctly specified for $E(Y|W=0,X=x)$. \end{thm}
The property indicated by Theorem~\ref{doublerobustthm} is known
as double robustness \cite{robins1994estimation}. Since double robustness
is a special case for multiple robustness which we discuss in Section \ref{sec:Multiple Robustness}, the proof for double robustness
is shown in the supplementary material where we prove the multiple robustness. Furthermore,
$\hat{\theta}_{\text{mis}}$ is asymptotically normal distributed if both $\pi_{1}(x;\alpha_{1})$
and $\pi_{0}(x;\alpha_{0})$ are correctly specified. The asymptotic
distribution for $\hat{\theta}_{\text{mis}}$ is shown in the next section, where
we describe our method in a more general way by allowing multiple models
for each of $\pi_{1}(x)$, $\pi_{0}(x)$, $E(Y|W=1, X=x)$ and $E(Y|W=0, X=x)$, but not only one model for each.

For comparison, we consider an alternative estimator for $\theta$,
which is $\widetilde{\theta}_{\text{qz}}=\sum_{i=1}^{m_{0}}\widetilde{p}_{i}Y_{1i}-\sum_{j=1}^{n_{0}}\widetilde{q}_{j}Y_{0j}$
where $\widetilde{p}_{i}$'s and $\widetilde{q}_{j}$'s are obtained
by the same optimization problem mentioned above except that the constraints
(\ref{eq:mis1}) - (\ref{eq:mis4}) are replaced by
\begin{eqnarray}
\sum_{i=1}^{m_{0}}p_{i}\pi_{1}(X_{1i};\hat{\alpha}_{1}) & = & \frac{1}{m}\sum_{i=1}^{m}\pi_{1}(X_{1i};\hat{\alpha}_{1})\label{eq:mis5},\\
\sum_{j=1}^{n_{0}}q_{j}\pi_{0}(X_{0j};\hat{\alpha}_{0}) & = & \frac{1}{n}\sum_{j=1}^{n}\pi_{0}(X_{0j};\hat{\alpha}_{0})\label{eq:mis6},\\
\sum_{i=1}^{m_{0}}p_{i}g(X_{1i}) & = & \frac{1}{m}\sum_{i=1}^{m}g(X_{1i})\label{eq:mis7},\\
\sum_{j=1}^{n_{0}}q_{j}h(X_{0j}) & = & \frac{1}{n}\sum_{j=1}^{n}h(X_{0j})\label{eq:mis8}.
\end{eqnarray}

Clearly, $\widetilde{\theta}_{\text{qz}}$ is obtained by directly applying the method
proposed in Qin and Zhang  \cite{qin2007empirical} which is originally
designed for the one-sample case. Since our method considers the randomization procedure
for the two samples in each constraint of (\ref{eq:mis1}) - (\ref{eq:mis4}),
while each one of (\ref{eq:mis5}) - (\ref{eq:mis8}) only focuses on the information from one of the two samples, $\hat{\theta}_{\text{mis}}$
is intuitively more efficient, which is confirmed by our simulation
results in Section \ref{sec:simu}.

\subsection{Multiple robustness}
\label{sec:Multiple Robustness}
Following the work of Han and Wang \cite{han2013estimation}
and Han \cite{han2014further}, we postulate multiple working
parametric models $\mathcal{P}_{1}=\{\pi_{1}^{c}(x;\alpha_{1}^{c});c=1,\ldots,C\}$
for $\pi_{1}(x)$, $\mathcal{P}_{0}=\{\pi_{0}^{d}(x;\alpha_{0}^{d});d=1,\ldots,D\}$
for $\pi_{0}(x)$, $\mathcal{G}=\{\widetilde{g}^{e}(x;\beta_{1}^{e});e=1,\ldots,E\}$
for $E(Y|W=1,X=x)$ and $\mathcal{H}=\{\widetilde{h}^{f}(x;\beta_{0}^{f});f=1,\ldots,F\}$
for $E(Y|W=0,X=x)$. The $\alpha_{1}^{c}$, $\alpha_{0}^{d}$, $\beta_{1}^{e}$,
$\beta_{0}^{f}$ are the corresponding parameters and their estimators are denoted as $\hat{\alpha}_{1}^{c}$,
$\hat{\alpha}_{0}^{d}$, $\hat{\beta}_{1}^{e}$, $\hat{\beta}_{0}^{f}$. Usually, the estimators of the postulated propensity score models, i.e.,   $\hat{\alpha}_{1}^{c}$ for $c=1\cdots C$, and  $\hat{\alpha}_{0}^{d}$ for $d=1\cdots D$, are taken to be the maximizer of the corresponding binomial likelihoods
\begin{eqnarray}
\prod_{i=1}^{m}\left\{ \pi_{1}^{c}\left(\alpha_{1}^{c};X_{1i}\right)\right\} ^{R_{1i}}\left\{ 1-\pi_{1}^{c}\left(\alpha_{1}^{c};X_{1i}\right)\right\} ^{1-R_{1i}}\label{eq:biolik1},\quad c=1,\ldots,C,\\
\prod_{j=1}^{n}\left\{ \pi_{0}^{d}\left(\alpha_{0}^{d};X_{0j}\right)\right\} ^{R_{0j}}\left\{ 1-\pi_{0}^{d}\left(\alpha_{0}^{d};X_{0j}\right)\right\} ^{1-R_{0j}}\label{eq:biolik0},\quad d=1,\ldots,D.
\end{eqnarray}
Our proposed estimator is
 $\hat{\theta}_{\text{mr}}=\sum_{i=1}^{m_{0}}{\hat{p}_{i}Y_{1i}}-\sum_{j=1}^{n_{0}}{\hat{q}_{j}Y_{0j}}$ with the estimated weights $\{(\hat{p}_{i},\hat{q}_{j});i=1,\cdots,m_{0},j=1,\cdots,n_{0}\}$ obtained by maximizing the empirical likelihood $\prod_{i=1}^{m_{0}}p_{i}\prod_{j=1}^{n_{0}}q_{j}$ in (\ref{eq:nonlik2}) with the same constraints except that (\ref{eq:mis1}) - (\ref{eq:mis4}) are changed to 
\begin{eqnarray}
\sum_{i=1}^{m_{0}}p_{i}\pi_{1}^{c}(X_{1i};\hat{\alpha}_{1}^{c}) & = & \overline{\pi}_{1}^{c}\ \ (c=1,\cdots,C),\label{eq:mis9}\\
\sum_{j=1}^{n_{0}}q_{j}\pi_{0}^{d}(X_{0j};\hat{\alpha}_{0}^{d}) & = & \overline{\pi}_{0}^{d}\ \ (d=1,\cdots,D),\label{eq:mis10}\\
\sum_{i=1}^{m_{0}}p_{i}g^{e}(X_{1i};\hat{\beta}_{1}^{e}) & = & \overline{g}^{e}\ \ (e=1,\cdots,E),\label{eq:mis11}\\
\sum_{j=1}^{n_{0}}q_{j}h^{f}(X_{0j};\hat{\beta}_{0}^{f}) & = & \overline{h}^{f}\ \ (f=1,\cdots,F)\label{eq:mis12},
\end{eqnarray}
where $\overline{\pi}_{1}^{c}=\frac{1}{N}\left\{ \sum_{i=1}^{m}\pi_{1}^{c}(X_{1i};\hat{\alpha}_{1}^{c})+\sum_{j=1}^{n}\pi_{1}^{c}(X_{0j};\hat{\alpha}_{1}^{c})\right\}$,  $\overline{\pi}_{0}^{d}= \frac{1}{N}\left\{\sum_{i=1}^{m}\pi_{0}^{d}(X_{1i};\hat{\alpha}_{0}^{d})  +\right.\left.\sum_{j=1}^{n}\pi_{0}^{d}(X_{0j};\hat{\alpha}_{0}^{d})\right\} $,
$\overline{g}^{e}=\frac{1}{N}\left\{
\sum_{i=1}^{m}g^{e}(X_{1i};\hat{\beta}_{1}^{e})
+\sum_{j=1}^{n}g^{e}(X_{0j};\hat{\beta}_{1}^{e})\right\} $, $\overline{h}^{f}=\frac{1}{N}\left\{ \sum_{i=1}^{m}h^{f}(X_{1i};\hat{\beta}_{0}^{f})+\right.\left.\sum_{j=1}^{n}h^{f}(X_{0j};\hat{\beta}_{0}^{f})\right\} $ with $c=1,\cdots,C$, $d=1,\cdots,D$, $e=1,\cdots,E$, $f=1,\cdots,F$.
The first two constraints ensure that $p_{i}$'s and $q_{j}$'s are
empirical probabilities as mentioned in Section \ref{sec:mis}. The latter four
constraints calibrate the weighted average of each postulated parametric
function, which is evaluated at one biased sample with missing outcomes,
to the corresponding empirical average of the two entire samples,
which consistently estimates the population mean. Unlike the previous
setting in Section \ref{sec:mis}, there are more than one postulated models
for each one of $\pi_{1}(x)$, $\pi_{0}(x)$, $E(Y|W=1,X=x)$ and $E(Y|W=0,X=x)$
to incorporate information from covariates. In this case, we have the following theorem on the consistency of $\hat{\theta}_{\text{mr}}$.
\begin{thm} \label{multiplerobustthm} $\hat{\theta}_{\text{mr}}$ is consistent
	for $\theta$ as $N\xrightarrow{}\infty$ if the following two conditions are satisfied: i) $\mathcal{P}_{1}$
	contains a correctly specified model for $\pi_{1}(x)$ or $\mathcal{G}$
	contains a correctly specified model for $E(Y|W=1,X=x)$; ii) $\mathcal{P}_{0}$
	contains a correctly specified model for $\pi_{0}(x)$ or $\mathcal{H}$
	contains a correctly specified model for $E(Y|W=0,X=x)$. \end{thm}
Therefore, $\hat{\theta}_{\text{mr}}$ is a multiple robust estimator of $\theta$.
Next, we introduce the asymptotic distribution and efficiency of $\hat{\theta}_{\text{mr}}$. The following theorem gives the asymptotic distribution of $\hat{\theta}_{\text{mr}}$.
\begin{thm} \label{misIF} When $\pi_{1}^{1}(x;\alpha_{1}^{1})$
	is a correctly specified model for $\pi_{1}(x)$ and $\pi_{0}^{1}(x;\alpha_{0}^{1})$
	is a correctly specified model for $\pi_{0}(x)$, $N^{1/2}(\hat{\theta}_{\text{mr}}-\theta)$
	is asymptotically normal distributed with mean $0$ and variance $\text{var}\{\varphi(Y,X,R,W)\}$
	with the influence function for $\hat{\theta}$
	\begin{align*}
	\varphi(Y,X,R,W)=&Z_{1}-Z_{0}-\frac{W}{\delta}E(Z_{1}S_{1}^{\top})\{E(S_{1}^{\otimes2})\}^{-1}S_{1}
	-\frac{1-W}{1-\delta}E(Z_{0}S_{0}^{\top})\{E(S_{0}^{\otimes2})\}^{-1}S_{0},	\end{align*}
	where
	\begin{align*}
	Z_{1}= & \frac{W}{\delta}\frac{R}{\pi_{1}\left(X\right)}(Y-\mu_{1})-\frac{WR-\delta\pi_{1}\left(X\right)}{\delta\pi_{1}\left(X\right)}L_{1}^{\top}G_{1}^{-1}U_{1}(X),\\
	Z_{0}= & \frac{1-W}{1-\delta}\frac{R}{\pi_{0}\left(X\right)}(Y-\mu_{0})
	-\frac{WR-\delta\pi_{0}\left(X\right)}{\delta\pi_{0}\left(X\right)}L_{0}^{\top}G_{0}^{-1}U_{0}(X),\\
	S_{1}\left(X_{1},R_{1},\alpha_{1}^{1}\right)= & \frac{R_{1}-\pi_{1}^{1}\left(\alpha_{1}^{1};X_{1}\right)}{\pi_{1}^{1}\left(\alpha_{1}^{1};X_{1}\right)\left\{ 1-\pi_{1}^{1}\left(\alpha_{1}^{1};X_{1}\right)\right\} }\frac{\partial\pi_{1}^{1}\left(\alpha_{1}^{1};X_{1}\right)}{\partial\alpha_{1}^{1}},\\
	S_{0}\left(X_{0},R_{0},\alpha_{0}^{1}\right)= & \frac{R_{0}-\pi_{0}^{1}\left(\alpha_{0}^{1};X_{0}\right)}{\pi_{0}^{1}\left(\alpha_{1}^{1};X_{0}\right)\left\{ 1-\pi_{0}^{1}\left(\alpha_{0}^{1};X_{0}\right)\right\} }\frac{\partial\pi_{0}^{1}\left(\alpha_{0}^{1};X_{0}\right)}{\partial\alpha_{0}^{1}}.
	\end{align*}
	Here, $S_{1}\left(X_{1},R_{1},\alpha_{1}^{1}\right)$ and $S_{0}\left(X_{0},R_{0},\alpha_{0}^{1}\right)$
	are the corresponding score functions of the binomial likelihoods in (\ref{eq:biolik1})
	and (\ref{eq:biolik0}), respectively.
\end{thm}

To show that our proposed estimator $\hat{\theta}_{\text{mr}}$ attains the semiparametric efficiency bound, we derive the semiparametric efficiency bound for ATE estimator in RCTs with missing outcomes, which is given by the following theorem.

\noindent \begin{thm} \label{misbound} The efficient influence function for the RAL estimators of $\theta$ in RCTs with missing outcomes is given by
	\begin{align*}
	\varphi_{opt}(Y,X,R,W)= & \frac{WR}{\delta\pi_{1}(X)}\left\{ Y-E(Y|W=1,X)\right\}
	-\frac{(1-W)R}{(1-\delta)\pi_{0}(X)}\left\{ Y-E(Y|W=0,X)\right\} \\
	& +E(Y|W=1,X)-E(Y|W=0,X)-\theta,
	\end{align*}
	which leads to the semiparametric efficiency bound  $\text{var}\left\{ \varphi_{opt}(Y,X,R,W)\right\}$.
\end{thm}

Following the techniques used in Han and Wang  \cite{han2013estimation},
we prove that the asymptotic variance $\text{var}\{\varphi(Y,X,R,W)\}$
in Theorem \ref{misIF} can reach the semiparametric efficiency bound
defined in Theorem \ref{misbound}, which leads to the following result
on the efficiency of $\hat{\theta}_{\text{mr}}$ (proofs are given in the supplementary material).
\begin{thm} When $\mathcal{P}_{1}$ contains a correctly specified
	model for $\pi_{1}(x)$, $\mathcal{P}_{0}$ contains a correctly specified
	model for $\pi_{0}(x)$, $\mathcal{G}$ contains a correctly specified
	model for $E(Y|W=1,X=x)$ and $\mathcal{H}$ contains a correctly
	specified model for $E(Y|W=0,X=x)$, the asymptotic variance of $\hat{\theta}_{\text{mr}}$
	attains the semiparametric efficiency bound. \end{thm} For comparison,
the alternative estimator, which is based on the work of Han and Wang
\cite{han2013estimation}, is denoted as  $\widetilde{\theta}_{\text{hw}}=\sum_{i=1}^{m_{0}}\widetilde{p}_{i}Y_{1i}-\sum_{j=1}^{n_{0}}\widetilde{q}_{j}Y_{0j}$
where $\widetilde{p}_{i}$'s and $\widetilde{q}_{j}$'s are obtained
from the same optimization problem with the same constraints as in Section \ref{sec:mis} except
that (\ref{eq:mis9}) - (\ref{eq:mis12}) are replaced by
\begin{eqnarray}
\sum_{i=1}^{m_{0}}p_{i}\pi_{1}^{c}(X_{1i};\hat{\alpha}_{1}^{c}) & =\frac{1}{m}\sum_{i=1}^{m}\pi_{1}^{c}(X_{1i};\hat{\alpha}_{1}^{c})\ \ (c=1,\cdots,C),\label{eq:mis13}\\
\sum_{j=1}^{n_{0}}q_{j}\pi_{0}^{d}(X_{0j};\hat{\alpha}_{0}^{d}) & =\frac{1}{n}\sum_{j=1}^{n}\pi_{0}^{d}(X_{0j};\hat{\alpha}_{0}^{d})\ \ (d=1,\cdots,D),\label{eq:mis14}\\
\sum_{i=1}^{m_{0}}p_{i}g^{e}(X_{1i};\hat{\beta}_{1}^{e}) & =\frac{1}{m}\sum_{i=1}^{m}g^{e}(X_{1i};\hat{\beta}_{1}^{e})\ \ (e=1,\cdots,E),\label{eq:mis15}\\
\sum_{j=1}^{n_{0}}q_{j}h^{f}(X_{0j};\hat{\beta}_{0}^{f}) & =\frac{1}{n}\sum_{j=1}^{n}h^{f}(X_{0j};\hat{\beta}_{0}^{f})\ \ (f=1,\cdots,F).\label{eq:mis16}
\end{eqnarray}

As mentioned in Section \ref{sec:mis}, our method takes randomization for the two
samples into account, as indicated by each of the constraints (\ref{eq:mis9})
- (\ref{eq:mis12}), while each of the constraint (\ref{eq:mis13})
- (\ref{eq:mis16}) only involves one of the two samples. Therefore,
$\hat{\theta}_{\text{mr}}$ is intuitively more efficient than $\widetilde{\theta}_{\text{hw}}$,
which is confirmed by our simulation results in Section \ref{sec:simu}.

\section{Optimization Details}
\label{sec:opt}
In Section \ref{sec:Numerical implementation}, we introduce
the computation details of solving the aforementioned optimization problem to obtain our proposed estimators, based on data with or without missing outcomes. Besides, we illustrate how to tackle the convex hull constraint problem in Section \ref{sec:Convex hull constraint problem}.

\subsection{Numerical implementation}
\label{sec:Numerical implementation}
As mentioned in Section \ref{sec:prop}, the proposed optimization problem actually can be split into two optimization problems to estimate $p_{j}$'s
and $q_{j}$'s separately. Now we demonstrate the method to estimate
$p_{j}$'s, the estimation of $q_{j}$'s  follows the same procedure. We
only need to maximize
\begin{equation}
\prod_{i=1}^{m}p_{i},\label{nonlik1}
\end{equation}
subject to $(\ref{eq:nomis1})$ and $(\ref{eq:nomis3})$. To
simplify the notation, we write $\widehat{U}_{1i}=g(X_{1i})-\bar{g}$. Applying
the standard Lagrange multiplier method, the solution of $p_{i}$ can be written as
\begin{equation}
  \label{eq:psolution}
\hat{p}_{i}=\frac{1}{m}\frac{1}{1+\hat{\lambda}_{1}^{\top}\widehat{U}_{1i}}
\end{equation}
where $\hat{\lambda}_{1}$ is the $r_1$-dimensional Lagrange multipliers satisfying
\begin{equation}
  \label{eq:lambdaderive}
\frac{1}{m}\sum_{i=1}^{m}\frac{\widehat{U}_{1i}}{1+\lambda_{1}^{{\top}}\widehat{U}_{1i}}=0.
\end{equation}
\noindent
In order to search for the solution of $\lambda_{1}$, we define
\begin{equation}
\label{eq:lambdasearch}
\tilde{l}(\lambda_1)=\sum_{i=1}^{m}\log\left(1+\lambda_{1}^{\top}\widehat{U}_{1i}\right)
\end{equation}
as our maximizer over $\lambda_{1}$,
which is a strictly convex function. The maximum point  $\hat{\lambda}_{1}$ of (\ref{eq:lambdasearch}) satisfies (\ref{eq:lambdaderive}) and the $\hat{p}_{i}$ given by (\ref{eq:psolution}) is subject to
(\ref{eq:nomis1}).
Note that the existence of the
solution of $\lambda_{1}$ requires some conditions including the convex hull constraint that the convex hull of $\{\widehat{U}_{1i}\}_{i=1}^{m}$ retains the
zero point. Here, we use a modified Newton–Raphson
algorithm to do the numerical
search for $\lambda_{1}$, which is similar to the method discussed by Chen et al. \cite{chen2002estimation}.

\begin{algorithm}

\caption{Modified Newton–Raphson algorithm}
  \label{alg:noaug}
Step 0. Let $\lambda_{1}^{(0)}=0$. Set $t=0$, $\gamma_{0}=1$ and $\varepsilon=10^{-8}$.

Step 1. Calculate $\Delta_{1}\left(\lambda_{1}^{(t)}\right)=\partial\tilde{l}/\partial\lambda_{1}$
and $\Delta_{2}\left(\lambda_{1}^{(t)}\right)=\left\{ \partial^{2}\tilde{l}/\left(\partial\lambda_{1}\partial\lambda_{1}^{\top}\right)\right\} ^{-1}\Delta_{1}\left(\lambda_{1}^{(t)}\right)$; that
is
\begin{align*}
\Delta_{1}(\lambda_{1})=&\sum_{i=1}^{m}\frac{\widehat{U}_{1i}}{1+\lambda_{1}^{\top}\widehat{U}_{1i}},\\
\quad\Delta_{2}(\lambda_{1})=&\left\{ -\sum_{i=1}^{m}\frac{\widehat{U}_{1i}\widehat{U}_{1i}^{\top}}{\left(1+\lambda_{1}^{\top}\widehat{U}_{1i}\right)^{2}}\right\} ^{-1}\Delta_{1}(\lambda_{1}).
\end{align*}
If $||\Delta_{2}\left(\lambda_{1}^{(t)}\right)||<\varepsilon$, stop
the algorithm and report $\lambda_{1}^{(t)}$; otherwise go to Step 1.

Step 2. Calculate $\delta^{(t)}=\gamma^{(t)}\Delta_{2}\left(\lambda_{1}^{(t)}\right).$
If $1+\left(\lambda_{1}^{(t)}-\delta^{(t)}\right)^{\top}\widehat{U}_{1i}\leqslant0$
for some $i$ or $\tilde{l}\left(\lambda_{1}^{(t)}-\delta^{(t)}\right)<\tilde{l}\left(\lambda_{1}^{(t)}\right)$,
let $\gamma^{(t)}=\gamma^{(t)}/2$ and repeat Step 2.

Step 3. Set $\lambda_{1}^{(t+1)}=\lambda_{1}^{(t)}-\delta^{(t)}$, $t=t+1$ and $\gamma^{(t+1)}=(t+1)^{-1/2}.$ Go to Step 1.
\end{algorithm}

Similarly, to obtain the $\hat{p}_{i}$'s by solving the optimization problems described in Section \ref{sec:mis} or \ref{sec:Multiple Robustness}, we only have to take $m=m_0$ in (\ref{nonlik1}) and define $\widehat{U}_{1i}=\{\pi_{1}(X_{1i};\hat{\alpha}_{1})-\overline{\pi}_{1},g(X_{1i},\hat{\beta}_{1})-\overline{g}\}^{{\top}}$
or $\widehat{U}_{1i}=\{\pi_{1}^{1}(X_{1i};\hat{\alpha}_{1}^{1})-\overline{\pi}_{1}^{1},...,\pi_{1}^{C}(X_{1i};\hat{\alpha}_{1}^{C})-\overline{\pi}_{1}^{C},g(X_{1i};\hat{\beta}_{1}^{1})-\overline{g}_{1}^{1},..., g(X_{1i};\hat{\beta}_{1}^{E})-\overline{g}_{1}^{E}\}^{{\top}}$, respectively.

\subsection{Convex hull constraint problem}
\label{sec:Convex hull constraint problem}
When we try to solve the constrained maximization problem
depicted in Section \ref{sec:prop},
a major problem encountered frequently in practise is that the convex hull condition, i.e., the zero vector is an interior point of the convex hull spanned by $\{\widehat{U}_{1i}\}_{i=1}^{m}$, may not be satisfied. The violation of the convex hull condition causes that
the solution for Lagrange
multipliers may not exist, leading to the non-convergence of the algorithm.

This convex hull constraint may be easily violated when the samples
are small or the constraints are high-dimensional. Some significant
efforts have been made to solve this problem. For instance, Emerson and Owen \cite{emerson2009calibration} proposed a balanced
augmented empirical likelihood (BAEL) method, which aims to augment
the sample with two artificial data points leading to an expanded
convex hull with the zero vector inside while preserving the mean of augmented
data as the same. Nguyen et al.\cite{nguyen2015simulation} extended Emerson and Owen's method \cite{emerson2009calibration} to the general estimating equations. Following their work, we define two
artificial points added in $\{\widehat{U}_{1i}\}_{i=1}^{m}$ as
\begin{eqnarray*}
 \widehat{U}_{1(m+1)} & = & -sc_{u}^{*}\bar{u}_1,\\
 \widehat{U}_{1(m+2)} & = & 2\overline{U}_1+sc_{u}^{*}\bar{u}_1,
\end{eqnarray*}
where $\overline{U}_1=\frac{1}{m}\sum_{i=1}^{m}\widehat{U}_{1i}$ is in
the direction of $\bar{u}_1=\frac{\overline{U}_1}{||\overline{U}_1||}$,
$c_{u_1}^{*}$ is defined as the inverse Mahalanobis distance of a unit
vector from $\overline{U}_1$ given by $c_{u_1}^{*}=(\bar{u}_1^{\top}S^{-1}\bar{u}_1)^{-1/2}$,
where $S$ is the sample covariance matrix, $s$ is an additional parameter set to tune the calibration of the resulting statistic. Note that the sample mean for $\widehat{U}_{1i}$ is maintained by adding these two points, i.e.,  $\frac{1}{m}\sum_{i=1}^m{\widehat{U}_{1i}}=\frac{1}{m+2}\sum_{i=1}^{m+2}{\widehat{U}_{1i}}=\overline{U}_1$.

After augmenting the sample as $\{\widehat{U}_{1i}\}_{i=1}^{m+2}$ and $\{\widehat{U}_{0j}\}_{j=1}^{n+2}$,
the empirical likelihood function for estimation of $\theta$ can
be adjusted as
\begin{equation*}
\prod_{i=1}^{m+2}p_{i}\prod_{j=1}^{n+2}q_{j}
\end{equation*}
subject to
\begin{align*}
\sum_{i=1}^{m+2}{p_{i}} & =1,\ \ p_{i}\geq0,\ \ i=1,\cdots,m, \\
\sum_{j=1}^{n+2}{q_{j}} & =1,\ \ q_{j}\geq0,\ \ j=1,\cdots,n, \\
\sum_{i=1}^{m+2}p_{i}\widehat{U}_{1i} & =0,\\
\sum_{j=1}^{n+2}q_{j}\widehat{U}_{0j} & =0.
\end{align*}

In this case, the solution for the weights is given by
\begin{equation*}
\hat{p}^{*}_{i}=\frac{1}{(m+2)}\frac{1}{(1+\hat{\lambda}_{1}^{*\top}\widehat{U}_{1i})},
\end{equation*}
and the $\hat{\lambda}^{*}_1$ is obtained by solving
\begin{equation*}
\frac{1}{m+2}\sum_{i=1}^{m+2}\frac{\widehat{U}_{1i}}{1+\lambda_{1}^{{\top}}\widehat{U}_{1i}}=0.
\end{equation*}
Then our maximizer over $\lambda_1$ changed to
\[
\widetilde{l}^{*}(\lambda_1)=\sum_{i=1}^{m+2}\log\left(1+\lambda_{1}^{\top}\widehat{U}_{1i}\right).
\]
Therefore, we provide another modified Newton–Raphson algorithm with an augmented sample in Algorithm \ref{alg:withaug} to avoid violation of the convex  hull constraint when searching for $\hat{\lambda}^{*}_1$. Since Algorithm~\ref{alg:withaug} only has one more step of generating two artificial points to build an augmented sample compared to Algorithm~\ref{alg:noaug}, these two algorithms have almost the same computational speed.

\begin{algorithm}
\caption{Modified Newton–Raphson algorithm with an augmented sample}
  \label{alg:withaug}
Step 0.
 Let $\lambda_{1}^{(0)}=0$. Set $t=0$, $\gamma_{0}=1$ and $\varepsilon=10^{-8}$.

Step 1. Generate two artificial points:
 \begin{eqnarray*}
	 \widehat{U}_{1(m+1)} & = & -sc_{u}^{*}\bar{u}_1,\\
	 \widehat{U}_{1(m+2)} & = & 2\overline{U}_1+sc_{u}^{*}\bar{u}_1.
 \end{eqnarray*}

Step 2. Calculate $\Delta_{1}\left(\lambda_{1}^{(t)}\right)=\partial\tilde{l}^{*}/\partial\lambda_{1}$
and $\Delta_{2}\left(\lambda_{1}^{(t)}\right)=\left\{ \partial^{2}\tilde{l}^{*}/\left(\partial\lambda_{1}\partial\lambda_{1}^{\top}\right)\right\} ^{-1}\Delta_{1}\left(\lambda_{1}^{(t)}\right)$, that
is
\begin{align*}
\Delta_{1}(\lambda_{1})=&\sum_{i=1}^{m+2}\frac{\widehat{U}_{1i}}{1+\lambda_{1}^{\top}\widehat{U}_{1i}},\\
\quad\Delta_{2}(\lambda_{1})=&\left\{ -\sum_{i=1}^{m+2}\frac{\widehat{U}_{1i}\widehat{U}_{1i}^{\top}}{\left(1+\lambda_{1}^{\top}\widehat{U}_{1i}\right)^{2}}\right\} ^{-1}\Delta_{1}(\lambda_{1}).
\end{align*}
If $||\Delta_{2}\left(\lambda_{1}^{(t)}\right)||<\varepsilon$, stop
the algorithm and report $\lambda_{1}^{(t)}$; otherwise go to Step 2.

Step 3. Calculate $\delta^{(t)}=\gamma^{(t)}\Delta_{2}\left(\lambda_{1}^{(t)}\right).$
If $1+\left(\lambda_{1}^{(t)}-\delta^{(t)}\right)^{\top}\widehat{U}_{1i}\leqslant0$
for some $i$ or $\tilde{l}^{*}\left(\lambda_{1}^{(t)}-\delta^{(t)}\right)<\tilde{l}^{*}\left(\lambda_{1}^{(t)}\right)$,
let $\gamma^{(t)}=\gamma^{(t)}/2$ and repeat Step 2.

Step 4. Set $\lambda_{1}^{(t+1)}=\lambda_{1}^{(t)}-\delta^{(t)}$, $t=t+1$ and $\gamma^{(t+1)}=(t+1)^{-1/2}.$ Go to Step 2.
\end{algorithm}

In the simulations implemented in Section \ref{sec:Simulations}, we use Algorithm~\ref{alg:withaug} only in the Simulation 3 where we apply our method on the simulated missing data by solving the optimization problem in Section \ref{sec:mis}. Recall that this optimization problem has two more moment constraints involving propensity score models, which can easily cause a high-dimension problem especially when we take the functions $g(x)$ and $h(x)$ as the identity functions.

\FloatBarrier

\section{Simulation and Real Data Analysis}
\label{sec:simu}
In this section, we report the results of several simulation experiments and a real data analysis for ACTG175 data to evaluate the performance of our proposed estimators.

\subsection{Simulation}
\label{sec:Simulations}
We present four simulation studies to demonstrate the performance of our proposed method based on 1000 Monte Carlo data sets.

\noindent
\textbf{Simulation 1.}
Similar to the simulation studies reported by Tsiatis et al.\cite{tsiatis2008covariate}, we conduct a simulation experiment based on ACTG175 data analysis in Section \ref{sec:Real Data analysis}.
In each simulated data set, we generate five continuous baseline covariates $(X_1, X_2, X_3, X_4, X_5)$ from a multivariate normal distribution with empirical mean and covariance matrix of the same variables in the ACTG175 data.
Besides, we generate each binary covariate in $(X_6, X_7, X_8, X_9, X_{10}, X_{11})$ from an independent Bernoulli distribution with their own data proportion in the ACTG175 data as parameters.
Independent of all the other variables, the treatment indicator $W$ is derived from Bernoulli($\delta$) with $\delta$ as the treatment assignment probability.
Finally, according to the covariates and the treatment assignment, the outcome variable CD4 count at 20 $\pm$ 5 weeks is generated from a normal distribution with the conditional mean (\ref{eq:aplilinear}) and conditional variance given after (\ref{eq:aplilinear}).

In each data set, we use our proposed method and the competing methods mentioned in Tsiatis et al.\cite{tsiatis2008covariate} to estimate $\theta$,  including ``Unadjusted'' estimator $\overline{Y}_1-\overline{Y}_0$, ``Change score'' estimator $\overline{Y}_1-\overline{Y}_0-(\overline{X}_1-\overline{X}_0)$, two semiparametric estimators proposed by Tsiatis et al.\cite{tsiatis2008covariate} with variable selection procedure ``Forward-1'' and ``Forward-2'' estimators, and two classical estimators
``ANCOVA''  estimator\cite{lesaffre2003note} and
``KOCH'' estimator\cite{koch1998issues}. Details for these competing estimators are shown in the supplementary material.

Table \ref{TableACTG175} shows the results of two cases: $N=2139$ and $\delta=0.75$; $N=400$ and
$\delta=0.5$. ELW-Identity and ELW-Linear are both our proposed two-sample ELW estimators. A ``benchmark'' estimator of $\theta$, which uses the true treatment-specific regression models, is also included for comparison. The former estimator takes $g(x)$ and $h(x)$ as identity functions, while the latter one sets $g(x)$ and $h(x)$ as linear regression functions that fitted separately by data from each treatment group. Table \ref{TableACTG175} shows that all adjusted estimators including our proposed ones have better performance in all evaluation metrics compared to the unadjusted estimator, e.g. they all have smaller bootstrap standard error, which implies covariate information incorporation can lead to an efficiency improvement. Furthermore, the result indicates our proposed ELW estimators can achieve a significant efficiency gain as they enjoy the smallest bootstrap standard error and mean square error among all estimates.


\begin{table}[htbp]
  \centering
  \caption{Results for simulation based on ACTG175 data}

\begin{threeparttable}
    \begin{tabular}{lcccc}
    \noalign{\global\arrayrulewidth=0.4mm}\hline
    Estimator & \multicolumn{1}{c}{Bias} &  \multicolumn{1}{l}{Ave.Boot.SE} &  \multicolumn{1}{c}{Cov.prob.boot.} & \multicolumn{1}{c}{MSE} \\
\noalign{\global\arrayrulewidth=0.3mm}\hline
$n=2139,\delta=0.75$&&&&\\
    Unadjusted & -0.127  & 6.736  & 0.955  & 43.942  \\
    Change scores & -0.155  & 5.627  & 0.954  & 30.368  \\
    Forward-1 & -0.157  & 5.159  & 0.954  & 25.139  \\
    Forward-2 & -0.112  & 5.281  & 0.961  & 25.574  \\
    ANCOVA & -0.175  & 5.179  & 0.954  & 25.331  \\
    KOCH  & -0.162  & 5.147  & 0.954  & 25.034  \\
    ELW-Identity & -0.141  & 5.146  & 0.957  & 25.001  \\
    ELW-Linear & -0.140  & 5.133  & 0.956  & 25.028  \\
    Benchmark & -0.139  & 5.113  & 0.954  & 24.850  \\
  &&&&\\
$n=400,\delta=0.5$&&&&\\

    Unadjusted & 0.004  & 13.756  & 0.939  & 202.402  \\
    Change scores & -0.563  & 11.665  & 0.954  & 132.685  \\
    Forward-1 & -0.439  & 10.985  & 0.948  & 121.313  \\
    Forward-2 & -0.412  & 14.409  & 0.962  & 124.378  \\
    ANCOVA & -0.533  & 10.939  & 0.950  & 120.614  \\
    KOCH  & -0.523  & 10.941  & 0.949  & 120.795  \\
    ELW-Identity & -0.344  & 10.971  & 0.945  & 120.466  \\
    ELW-Linear & -0.381  & 11.008  & 0.945  & 120.794  \\
    Benchmark & -0.353  & 10.801  & 0.949  & 115.672  \\
\noalign{\global\arrayrulewidth=0.4mm}\hline
    \end{tabular}
    \begin{tablenotes}[flushleft]
      \small
      \item Bias is the mean difference between the estimator between $\hat{\theta}$ and the true value of $\theta$; Ave.Boot.SE is the average bootstrap standard error calculated as the average of 1000 bootstrap standard error estimates, each of which involves 500 bootstrap replicates; Cov.prob.boot. is the coverage probability of a 95$\%$ Wald confidence interval using the average bootstrap standard error as standard error; MSE is the mean squared error calculated as the mean squared difference between $\hat{\theta}$ and the true value of $\theta$.
       Details for each competing estimator are shown in the supplementary material.
    \end{tablenotes}
    \end{threeparttable}
    \label{TableACTG175}
\end{table}

\noindent
\textbf{Simulation 2.}
The above simulation design assumes that there is a linear relationship between the outcome variable and covariates, which may not be true in most cases.
Next, we consider a nonlinear case to check the performance of our proposed method.
This simulation uses three continuous variables, $X=(X_{1},X_{2},X_{3})^{\top}\sim Normal(\mu,\Sigma)$, where
$\mu=(1,2,3)^{\top}$ and $\Sigma_{3\times3}= \left( \begin{matrix}
   1 & 1 & 1 \\
   1 & 2 & 2 \\
   1 & 2 & 3
  \end{matrix} \right)$.
We generate the outcome for each treatment group using $Y_{n}=\beta_{n0}^{(w)}+\beta_{n1}^{(w)}sin(X_{1})+\beta_{n2}^{(w)}X_{2}+\beta_{n3}^{(w)}X_{3}+\epsilon_{1}^{(w)}$,
where $w$ is the treatment assignment indicator that takes 1 for the treatment group and 0 for the control group. For comparison, we generate a similar linear outcome variable $Y_{l}$, where $Y_{l}=\beta_{l0}^{(w)}+\beta_{l1}^{(w)}X_{1}+\beta_{l2}^{(w)}X_{2}+\beta_{l3}^{(w)}X_{3}+\epsilon_{2}^{(w)}$, $w=0,1$. The only difference between the above two cases lies in the relationship between $Y$ and $X_{1}$. Let $(\epsilon_{1}^{(1)},\epsilon_{1}^{(0)},\epsilon_{2}^{(1)},\epsilon_{2}^{(0)})^{\top}\sim Normal(0,\Sigma_{\epsilon})$,
where $\Sigma_{\epsilon}$ is a diagonal matrix with diagonal entries $\{4^{2},6^{2},4^{2},6^{2}\}$. By setting $\beta_{n}^{(1)\top}=(12,11.756,10,9)$, $\beta_{n}^{(0)\top}=(9,19.593,13,10)$,
$\beta_{l}^{(1)\top}=(3,10,13,10)$ and $\beta_{l}^{(0)\top}=(5,7,10,9)$,
we control the true value of treatment effect $\theta$ between two treatment groups to be 10.  The sample size $N$ and the probability of treatment assignment $\delta$ for this
simulation are set to be $400$ and $0.5$.

\begin{table}[htbp]
  \centering
  \caption{Results for simulation comparing nonlinear and linear cases }
  \begin{threeparttable}
    \begin{tabular}{lcccc}
    \noalign{\global\arrayrulewidth=0.4mm} \hline
    \noalign{\global\arrayrulewidth=0.3mm}
    Estimator & \multicolumn{1}{c}{Bias} &  \multicolumn{1}{l}{Ave.Boot.SE} &  \multicolumn{1}{c}{Cov.prob.boot.} & \multicolumn{1}{c}{MSE} \\\hline
Nonlinear Case&&&&\\
    Unadjusted & 0.230  & 3.589  & 0.943  & 13.180  \\
    Forward-1 & 0.056  & 0.907  & 0.944  & 0.862  \\
    Forward-2 & 0.006  & 0.722  & 0.945  & 0.514  \\
    ANCOVA & 0.059  & 0.906  & 0.938  & 0.863  \\
    KOCH  & 0.057  & 0.905  & 0.940  & 0.861  \\
    Identity & 0.046  & 0.908  & 0.947  & 0.856  \\
    Linear model & 0.050  & 0.922  & 0.946  & 0.861  \\
    Benchmark & 0.016  & 0.660  & 0.949  & 0.433  \\

    &&&&\\
    Linear Case&&&&\\
        Unadjusted & 0.217  & 3.789  & 0.944  & 15.229  \\
        Forward-1 & 0.028  & 0.650  & 0.949  & 0.400  \\
        Forward-2 & 0.028  & 0.653  & 0.950  & 0.403  \\
        ANCOVA & 0.026  & 0.652  & 0.949  & 0.401  \\
        KOCH  & 0.028  & 0.651  & 0.953  & 0.400  \\
        ELW-Identity & 0.028  & 0.649  & 0.952  & 0.400  \\
        ELW-Linear & 0.029  & 0.651  & 0.950  & 0.400  \\
        Benchmark & 0.027  & 0.650  & 0.952  & 0.398  \\
    \noalign{\global\arrayrulewidth=0.4mm}
    \hline
    \end{tabular}%

  \begin{tablenotes}[flushleft]
    \small
    \item All entries are as in Table \ref{TableACTG175}.
  \end{tablenotes}
  \end{threeparttable}
  \label{Tablenonlinear}
\end{table}%

As shown in Table \ref{Tablenonlinear}, all estimators have better performance in the
linear case than in the nonlinear case, as we note that the mean squared error for each estimator in the nonlinear case is nearly twice of the mean squared error in the linear case except the unadjusted estimator and Forward-2 estimator.
Although all the estimators have very close results in the nonlinear case, which is indicated by the mean squared error, our proposed ELW estimators still achieve better precision than the others, but not as good as the Forward-2 estimator.

\noindent 
\textbf{Simulation 3.}
To evaluate the performance of our proposed estimator $\hat{\theta}_{\text{mis}}$
in Section \ref{sec:nomis}, which considers missing outcomes, we design a simulation
experiment to compare it with $\widetilde{\theta}_{\text{qz}}$, the estimator
proposed by Qin and Zhang \cite{qin2007empirical}. This simulation
involves four mutually independent variables, $X_{1}\sim Normal(1,3)$,
$X_{2}\sim Normal(2,3)$, $X_{3}\sim Normal(3,1)$ and $X_{4}\sim Bernoulli(0.5)$.
The outcome is generated by $Y_{m}=\beta_{m0}^{(w)}+\beta_{m1}^{(w)}X_{1}+\beta_{m2}^{(w)}X_{2}+\beta_{m3}^{(w)}X_{3}+\beta_{m4}^{(w)}X_{4}+\epsilon_{3}^{(w)}$,
$w=0,1$. We set $(\epsilon_{3}^{(0)},\epsilon_{3}^{(1)})^{\top}\sim Normal(0,\Sigma_{2})$,
and $\Sigma_{2}$ is a $2\times2$ diagonal matrix with the diagonal entries being $\{ 4^{2}, 6^{2} \} $. The true treatment effect is controlled to be 10 by setting
$\beta_{m}^{(0)\top}=(10,8,11,10,4)$ and $\beta_{m}^{(1)\top}=(5,7,10,9,6)$.
The missingness mechanism is set by logistic regression models $logit\{\pi_{w}(X,\alpha^{(w)})\}=\alpha^{(w)}_0+\alpha^{(w)}_1X_{1}+\alpha^{(w)}_2X_{2}+\alpha^{(w)}_3X_{3}+\alpha^{(w)}_4X_{4}$,
$w=0,1$. We use different set of $\alpha^{(w)}$ to change the missing
proportion of the outcomes. For example, we set $\alpha^{(1)}=(-5.147, -0.3, 0.8, 0.5, 0.3)^{\top}$ and $\alpha^{(0)}=(-3.247, 0.2, -0.3, 0.4, 0.5)^{\top}$ for a missing proportion of approximate $10\%$.

Table \ref{tablemis} reports the results of 1000 Monte Carlo data sets, in which we set $N=400$ and $\delta=0.5$. The bootstrap standard error in each Monte Carlo data set is based on 500 replicates. For each data set, we estimate $\theta$ using $\hat{\theta}_{\text{mis}}$ and $\widetilde{\theta}_{\text{qz}}$ for comparison. Results for estimators using the true model are included as the ``benchmark'' estimator. The evaluation metrics in Table \ref{tablemis} are the same as those in the previous experiments, noting that ``.qz'' indicates
this metric is for Qin and Zhang's method\cite{qin2007empirical}.

As shown in Table \ref{tablemis}, as the missing proportion increases, though all the estimators perform worse, our proposed estimators are still significantly better than Qin and Zhang's.
We note that $\hat{\theta}_{\text{mis}}$ and $\widetilde{\theta}_{qz}$ have close efficiency
judging from their close average bootstrap standard error and mean squared error when the missing proportion is low.
However, when the missing proportion is large, the performance of both $\hat{\theta}_{\text{mis}}$ and $\widetilde{\theta}_{\text{qz}}$ using identity functions deteriorates dramatically while those using a linear regression model, which is the correctly specified model, can maintain good performance.
This demonstrates a growing sensitivity to the model specified with a growing missing proportion
no matter using our proposed method or Qin and Zhang's.

\begin{table}[htbp]
  \centering
  \caption{Results for simulation with different missing proportion}
  \begin{threeparttable}
    \begin{tabular}{ccccccc}\hline
     \multicolumn{1}{c}{\multirow{2}{*}{Metric}} & \multirow{2}{*}{Estimator} &\multicolumn{5}{c}{Mean Missing Proportion}  \\
     \multicolumn{1}{c}{} &       & 0.138  & 0.242  & 0.333  & 0.417  & 0.501  \\
     \noalign{\global\arrayrulewidth=0.3mm}\hline

     \multirow{3}[0]{*}{Bias}  & Identity & -0.174  & -0.187  & -0.223  & -0.341  & -1.150  \\
          & Linear & -0.140  & -0.120  & -0.115  & -0.088  & -0.018  \\
          & Benchmark  & -0.140  & -0.123  & -0.111  & -0.090  & -0.022\\
    \noalign{\global\arrayrulewidth=0.2mm}\hline
    \multirow{3}[0]{*}{Bias.qz} & Identity & -0.170  & -0.186  & -0.226  & -0.387  & -1.186  \\
          & Linear & -0.139  & -0.119  & -0.109  & -0.090  & -0.013  \\
          & Benchmark  & -0.139  & -0.121  & -0.111  & -0.091  & -0.020  \\
\hline
    \multirow{3}[0]{*}{Ave.Boot.SE}  & Identity & 0.603  & 0.647  & 0.771  & 1.965  & 5.960  \\
          & Linear & 0.574  & 0.625  & 0.689  & 0.762  & 0.862  \\
          & Benchmark  & 0.576  & 0.625  & 0.687  & 0.760  & 0.857  \\
    \hline
    \multirow{3}[0]{*}{Ave.Boot.SE.qz} & Identity & 2.238  & 2.244  & 2.299  & 3.154  & 6.422  \\
          & Linear & 2.243  & 2.251  & 2.264  & 2.283  & 2.312  \\
          & Benchmark  & 2.247  & 2.253  & 2.266  & 2.283  & 2.311  \\
    \hline
    \multirow{3}[0]{*}{Cov.prob.boot} & Identity & 0.939  & 0.934  & 0.946  & 0.988  & 0.984  \\
          & Linear & 0.930  & 0.943  & 0.934  & 0.938  & 0.943  \\
          & Benchmark  & 0.928  & 0.945  & 0.937  & 0.945  & 0.944  \\
    \hline
    \multirow{3}[0]{*}{Cov.prob.boot.qz} & Identity & 0.957  & 0.958  & 0.958  & 0.977  & 0.981  \\
          & Linear & 0.954  & 0.958  & 0.954  & 0.955  & 0.958  \\
          & Benchmark  & 0.959  & 0.961  & 0.960  & 0.957  & 0.960  \\
    \hline
    \multirow{3}[0]{*}{MSE} & Identity & 0.383  & 0.443  & 0.546  & 1.054  & 18.427  \\
          & Linear & 0.368  & 0.418  & 0.514  & 0.581  & 0.743  \\
          & Benchmark  & 0.365  & 0.418  & 0.502  & 0.573  & 0.717  \\
    \hline
    \multirow{3}[0]{*}{MSE.qz} & Identity & 4.722  & 4.738  & 4.892  & 5.689  & 23.610  \\
          & Linear & 4.714  & 4.739  & 4.874  & 5.015  & 5.039  \\
          & Benchmark  & 4.711  & 4.734  & 4.899  & 5.009  & 5.005  \\
          \noalign{\global\arrayrulewidth=0.4mm}\hline
    \end{tabular}%

    \begin{tablenotes}[flushleft]
      \small
      \item All metrics are as in Table 1 except that metrics with no suffix are for our proposed estimator while those with ``.qz'' are for Qin and Zhang's method.  \end{tablenotes}
    \end{threeparttable}

  \label{tablemis}%
\end{table}%

\noindent
\textbf{Simulation 4.}
Table \ref{tablenomisMR} and Table \ref{tablemisMR} summarize the performance of $\hat{\theta}_{mr}$, which described in Section \ref{sec:Multiple Robustness} based on data with and without missing outcomes, respectively.


When considering data without missing outcomes, we estimate the ATE under a similar setting as in the last simulation. The outcome variable is generated by $Y_{c}=\beta_{c0}^{(w)}+\beta_{c1}^{(w)}X_{1}+\beta_{c2}^{(w)}X_{2}+\beta_{c3}^{(w)}X_{3}+\beta_{c4}^{(w)}X_{4}+\epsilon_{3}^{(w)}$, $w=0,1$.
The four mutually independent variables $X_1, X_2, X_3, X_4  $ are set to have the same distribution as in the last simulation. Here, we set $\beta_{c}^{(1)}=(10,10,0,0,0)^{\top}$
and $\beta_{c}^{(0)}=(3,7,0,0,0)^{\top}$, which lead to a true value of
$\theta=10$ and a true linear model only including $X_{1}$ to describe
the true relationship between outcome and covariates. In this way,
a series of identity functions used in the estimation can be regarded
as multiple models, one of which correctly specifies the true model,
as shown in the first row in Table \ref{tablenomisMR}.
The second row is related to another estimator using two linear regression models, each of which involves all 4 variables. The third estimator based on two linear regression models, both of which include only $X_{1}$, uses the exactly correct-specified model.
The results show a very close performance for these three estimators, which indicates the multiple robustness of the proposed estimator.

When we consider data with missing outcomes, we use a similar simulation setting as in Han \cite{han2014multiply}, which is originally designed to estimate the parameters in regression models. Denote four mutually independent covariates to be
$X_{1}\sim$$Normal(5,1)$, $X_{2}\sim$ $Bernoulli(0.5)$, $X_{3}\sim$$Normal(0,1)$
and $X_{4}\sim$ $Normal(0,1)$. The outcome is generated by $Y_{r}=\beta_{r0}^{(w)}+\beta_{r1}^{(w)}X_{1}+\beta_{r2}^{(w)}X_{2}+\beta_{r3}^{(w)}X_{3}+\beta_{r4}^{(w)}X_{4}+\epsilon_{Y}^{(w)}, w=0,1$,
where $\beta_{r}^{(1)\top}=(10,8,12,10,4)$ and $\beta_{r}^{(0)\top}=(6,7,10,9,6)$ leading to a true value of $\theta=10$. There are three auxiliary variables
involved: $S_{1}=1+X_{1}-X_{2}+$$\epsilon_{1},~ S_{2}=\mathcal{I}\left\{ S_{1}+0.3\epsilon_{2}>5.8\right\}$,
and $S_{3}=\exp\left[\left\{ S_{1}/9\right\} ^{2}\right]+$ $\epsilon_{3}.$
Here, $\mathcal{I}(\cdot)$ represents the indicator function, $\left(\epsilon_{Y},\epsilon_{1},\epsilon_{2},\epsilon_{3}\right)^{\top}\sim$
$Normal(\mathbf{0},\Sigma)$ where $\Sigma$ is a $4\times4$ matrix with diagonal entries $2,2,1$ and $1$, $(1,2)$-entry and $(2,1)$-entry
$0.5,$ and all the other entries 0. The missingness mechanism is
set by $logit\{\pi_{w}(X,S)\}=3.5-5.0S_{2}$, $w=0,1$, resulting in approximately
$37\%$ of missing outcome $Y_{r}$.

Following the above data setting, in addition to giving four correct models:
$\pi_{1}^{1}(X,\alpha_{1}^{1})=\alpha_{10}^{1}+\alpha_{11}^{1}S_{2}$,
$\pi_{0}^{1}(X,\alpha_{0}^{1})=\alpha_{00}^{1}+\alpha_{01}^{1}S_{2}$,  $g^{1}(X,\beta_{1}^{1})=\beta_{10}^{1}+\beta_{11}^{1}X_{1}+\beta_{12}^{1}X_{2}+\beta_{13}^{1}X_{3}+\beta_{14}^{1}X_{4}+\beta_{15}^{1}S_{1}$
and $h^{1}(X,\beta_{0}^{1})=\beta_{00}^{1}+\beta_{01}^{1}X_{1}+\beta_{02}^{1}X_{2}+\beta_{03}^{1}X_{3}+\beta_{04}^{1}X_{4}+\beta_{05}^{1}S_{1}$,
we also define an incorrect model for each model as
$\pi_{1}^{2}(X,\alpha_{1}^{2})=\alpha_{10}^{2}+\alpha_{11}^{2}X_1+\alpha_{12}^{2}X_2+\alpha_{13}^{2}X_3+
\alpha_{14}^{2}X_4+\alpha_{15}+S_{1}$,
$\pi_{0}^{2}(X,\alpha_{0}^{2})=\alpha_{00}^{2}+\alpha_{01}^{2}X_1+\alpha_{02}^{2}X_2+\alpha_{03}^{2}X_3+
\alpha_{04}^{2}X_4+\alpha_{05}+S_{1}$,
 $g^{2}(X,\beta_{1}^{2})=\beta_{10}^{2}+\beta_{11}^{2}S_{1}+\beta_{12}^{2}S_{2}+\beta_{13}^{2}S_{3}$
and $h^{2}(X,\beta_{0}^{2})=\beta_{00}^{2}+\beta_{01}^{2}S_{1}+\beta_{02}^{2}S_{2}+\beta_{03}^{2}S_{3}$
to test the multiple robustness of our proposed estimator.

From now on, all the eight models are used to estimate $\theta$ in the optimization problem with the constraints depicted in Section \ref{sec:Multiple Robustness}.
We consider the sample size to be $N = 400$, and the results are summarized based on 1000 replications.
In order to distinguish the estimators of different models, we assign a name for each in the form of ``ELW-00000000'', where the eight digits, from left to right,
indicate whether $\pi_{1}^{1}(X,\alpha_{1}^{1})$, $\pi_{1}^{2}(X,\alpha_{1}^{2})$,
$g^{1}(X,\beta_{1}^{1})$, $g^{2}(X,\beta_{1}^{2})$,  $\pi_{0}^{1}(X,\alpha_{0}^{1})$,
$\pi_{0}^{2}(X,\alpha_{0}^{2})$, $h^{1}(X,\beta_{0}^{1})$ or $h^{2}(X,\beta_{0}^{2})$ has been used in the estimation, by assigning 0 or 1 to the corresponding digit.

For implementation, $\widetilde{\theta}_{\text{hw}}$ is obtained by using
R-package \texttt{MultiRobust}, where we subtract two mean estimators for
the two samples by implementing the MR.mean function. Our proposed estimators are obtained by applying Algorithm \ref{alg:noaug}. According to the results in Table \ref{tablemisMR}, the
multiple robustness for all the estimators except ``ELW-01010101''
is well demonstrated since they all have ignorable bias. The efficiency performance of our proposed estimators
are consistently better than $\widetilde{\theta}_{\text{hw}}$. We find
that the estimators of ``ELW-10111011''
and ``ELW-11101110'' already have very
similar efficiency performance compared to ``ELW-10101010''
estimator where all the models are correctly specified.

\begin{table}[htbp]
  \centering
  \caption{Results for multiple robustness given data without missing outcomes}
    \begin{tabular}{lcccc}
      \noalign{\global\arrayrulewidth=0.3mm}\hline
    Estimator & \multicolumn{1}{l}{Bias} &  \multicolumn{1}{l}{Ave.BootSE} &  \multicolumn{1}{l}{Cov.prob.boot} & \multicolumn{1}{l}{MSE} \\
    \noalign{\global\arrayrulewidth=0.3mm}\hline
    Identity & -0.029   & 0.573    & 0.937  & 0.331  \\
    Linear & -0.029    & 0.572    & 0.944  & 0.330  \\
    Linear(correct) & -0.029  & 0.570  & 0.945  & 0.331  \\
    \noalign{\global\arrayrulewidth=0.3mm}\hline
    \end{tabular} %
  \label{tablenomisMR} %
\end{table}%

\begin{table}[htbp]
  \centering
  \caption{Results for multiple robustness given data with missing outcomes}
  \begin{threeparttable}
    \begin{tabular}{lcccc}\noalign{\global\arrayrulewidth=0.4mm}\hline
    Estimator & \multicolumn{1}{l}{Bias} & \multicolumn{1}{l}{Bias.hw} & \multicolumn{1}{l}{MSE} & \multicolumn{1}{l}{MSE.hw}
    \\\noalign{\global\arrayrulewidth=0.3mm}\hline
    ELW-10101010 & -0.007  & 0.033  & 0.087  & 2.261  \\
    ELW-01010101 & 0.110  & 0.134  & 6.966  & 6.912  \\
    ELW-11111111 & -0.009  & 0.033  & 0.090  & 2.266  \\

\hline
ELW-10011001 & 0.119  & 0.122  & 6.693  & 6.576  \\
ELW-10101001 & 0.006  & 0.030  & 2.554  & 4.009  \\
ELW-10011010 & 0.106  & 0.126  & 3.169  & 4.689  \\
ELW-10111011 & -0.008  & 0.034  & 0.090  & 2.260  \\

\hline
ELW-01100110 & 0.003  & 0.043  & 0.088  & 2.260  \\
ELW-10100110 & -0.002  & 0.038  & 0.090  & 2.272  \\
ELW-01101010 & -0.001  & 0.039  & 0.085  & 2.250  \\
ELW-11101110 & 0.003  & 0.043  & 0.088  & 2.260  \\

\noalign{\global\arrayrulewidth=0.4mm}\hline
    \end{tabular}%
    \begin{tablenotes}[flushleft]
      \small
      \item All metrics are as in Table 1 except that metrics with no suffix are for our proposed estimator while those with ``.hw'' are for Han and Wang's method.
    \end{tablenotes}
    \end{threeparttable}
  \label{tablemisMR}%
\end{table}%

\FloatBarrier

\subsection{Real data analysis}
\label{sec:Real Data analysis}

Firstly, we demonstrate and compare our proposed method with the other 5 competing methods by applying all of them to ACTG 175 data, which is collected from 2139 HIV-infected individuals and equally randomizes all of them to 4 different antiretroviral regimens: zidovudine (ZDV) monotherapy, ZDV + didanosine (ddI), ZDV + zalcitabine, and ddI monotherapy.

Simplifying the experiment setting as Tsiatis et al.\cite{tsiatis2008covariate} did, we regard the $m=532$ individuals receiving ZDV monotherapy as the treatment group, while the rest of $n =1607$ individuals receiving any other antiretroviral regimens were classified as the control group. Accordingly, we have $\delta = \frac{m}{m+n} \approx 0.75$.

We focus on the analysis of mean differences in CD4 count (cells/mm$^{3}$) at 20 $\pm$ 5 weeks post-baseline (CD420), denoted as $Y$, between the above 2 groups.
For potential use in covariate adjustment, we consider the following 5 continuous baseline variables: $X_1=$CD4 count (cells/mm$^{3}$), $X_2=$CD8 count (cells/mm$^{3}$), $X_3=$age(years), $X_4=$weight (kg), $X_5=$Karnofsky score (scale of 0–100), and 7 indicator variables: $X_6=$hemophilia, $X_7=$homosexual activity, $X_8=$history of intravenous drug use, $X_9=$race  (0=white, 1=nonwhite), $X_{10}=$gender (0=female, 1=male), $X_{11}=$antiretroviral history (0=naive, 1=experienced), and $X_{12}=$symptomatic status (0=asymptomatic, 1=symptomatic).

Now we apply the optimization algorithm in Section \ref{sec:opt} to obtain the proposed ELW estimators.
We assume $g(X)$ and $h(X)$ to be linear regression functions or identity functions of covariates in two different scenarios.
In the first scenario, we develop two treatment-specific linear models for $E(Y|W=w,X)$, $w=0, 1$, with $12$ baseline covariates by fitting separate linear models to the observed data in each treatment arm. The fitted treatment-specific linear regression models are
\begin{equation}
\label{eq:aplilinear}
\begin{aligned}
g(X;\hat{\beta}_1)=&98.900+0.689X_1-0.019X_2-0.362X_3+0.133X_4\\
&+1.107X_5-17.337X_6+6.542X_7+12.026X_8\\
&-23.343X_9-13.301X_{10}-40.456X_{11}-20.545X_{12},\\
h(X;\hat{\beta}_0)=&126.771+0.719X_1-0.022X_2-0.432X_3-0.455X_4\\
&+0.607X_5-58.747X_6-19.672X_7-10.567X_8\\
&-5.818X_9+18.900X_{10}-41.816X_{11}-11.039X_{12}
\end{aligned}
\end{equation}
\noindent
with estimated treatment-specific variances $\widehat{Var}(Y|W=1,X)=(96.305)^{2}$
and $\widehat{Var}(Y|W=0,X)=(116.864)^{2}$, and the
treatment-specific coefficients of determination $R^{2}=0.3687$ for $W=1$ and $R^{2}=0.4592$ for $W=0$. Applying these models to the optimization procedure proposed in Section \ref{sec:prop}, we obtain the proposed  ELW-Linear estimator. In the second scenario, we replace linear functions with identity functions in the above models to obtain the ELW-Identity estimator. Here, $X$ denote the $l\times 1$ covariate vector with $l=12$.

\begin{table}[htbp]
\centering
\caption{Estimate of $\theta$ for the ACTG 175 data based on CD420}
\begin{threeparttable}
  \begin{tabular}{lcccc}
  \noalign{\global\arrayrulewidth=0.4mm}\hline
    Estimator    & \multicolumn{1}{c}{Estimate} &  \multicolumn{1}{c}{Boot.SE} & \multicolumn{1}{c}{Test stat.} & \multicolumn{1}{c}{Rel} \\\noalign{\global\arrayrulewidth=0.3mm}\hline
        Unadjusted & 46.810  & 7.055  & 6.924  & 1.000  \\
        Change scores & 50.409  & 5.693  & 9.150  & 1.506  \\
        Forward-1 & 49.895  & 5.439  & 9.716  & 1.733  \\
        Forward-2 & 51.589  & 5.700  & 10.183  & 1.780  \\
        ANCOVA & 49.694  & 5.451  & 9.680  & 1.734  \\
        KOCH  & 49.758  & 5.458  & 9.641  & 1.716  \\
        ELW-Identity & 50.006  & 5.288  & 10.057  & 1.849  \\
        ELW-Linear  & 49.824  & 5.200  & 9.776  & 1.760  \\
  \noalign{\global\arrayrulewidth=0.4mm}\hline
  \end{tabular}%
  \begin{tablenotes}[flushleft]
    \small
    \item Boot.SE is the boostrap-based
standard error; Test stat. is the Wald test statistic; and Rel. eff.
= (SE for the unadjusted estimator)$^{2}$/(SE for the indicated estimator)$^{2}$.
  \end{tablenotes}
  \end{threeparttable}
\label{tab:appliACTG1}%
\end{table}%

Given the results in Table \ref{tab:appliACTG1}, all different estimators indicate the same evidence of treatment difference. The performance of all methods seems to be similar except that the unadjusted estimator has a lower estimate due to a mild
imbalance for baseline CD4 between two treatment groups \cite{tsiatis2008covariate}. However, the bootstrap standard errors of our proposed ELW estimators are both smaller than that of the others, which indicates a better performance of our proposed method.

 \begin{table}
  \centering
  \caption{Estimate of $\theta$ for the ACTG 175 data based on CD496}
     \begin{threeparttable}
     \begin{tabular}{lccc}
     \noalign{\global\arrayrulewidth=0.4mm}\hline
     Estimator & \multicolumn{1}{l}{Estimate} & \multicolumn{1}{l}{Boot.SE} & \multicolumn{1}{l}{Test stat.}
     \\\noalign{\global\arrayrulewidth=0.3mm}\hline
     ELW-Identity & 64.623  & 9.082  & 7.116  \\
     ELW-Linear & 64.038  & 9.065  & 7.064  \\
     Qz-Identity & 61.223  & 10.316  & 5.935  \\
     Qz-Linear & 60.981  & 10.159  & 6.003  \\
     \noalign{\global\arrayrulewidth=0.4mm}\hline
     \end{tabular}%
     \begin{tablenotes}[flushleft]
      \small
      \item ELW-Identity and ELW-Linear are our proposed estimators using identity functions and linear functions, respectively. Similarly, Qz-Identity
 and Qz-Linear are the corresponding estimators based on Qin and Zhang's method\cite{qin2007empirical}.
     \end{tablenotes}
     \end{threeparttable}
  \label{tab:appliACTG2}%
 \end{table}%

Table \ref{tab:appliACTG2} shows the results for the estimates of $\theta$ based on the missing
outcome CD496, approximately 37$\%$ of which are missing.  Here, we only calculate the standard error using bootstrapping method. As shown in the Table \ref{tab:appliACTG2}, our proposed ELW estimators have higher
estimates of $\theta$ but consistently smaller bootstrap-based standard
errors than those based on Qin and Zhang's method\cite{qin2007empirical}, which indicates a
better efficiency for our proposed ELW estimators.
\FloatBarrier

\section{Conclusion}
\label{sec:concl}
We have proposed a two-sample empirical likelihood weighted estimator to effectively incorporate covariate information into the estimation of the average treatment effect in randomized clinical trials. Namely, we obtain two classes of estimated weights through constrained empirical likelihood estimation, where the constraints are designed to carry side information from covariates. Besides, our proposed estimator maintains objectivity since it separates the estimation of ATE from analysis of the covariate outcome relationship.

Furthermore, we apply the proposed estimator to the common problem of missing outcome data in RCTs under the assumption of missing at random. Theoretically, we have proved that our proposed estimator maintains double robustness and multiple robustness properties.

To evaluate the efficiency of our estimator, we demonstrate the proposed estimator is semiparametric efficient given data without or with missingness.
Various simulation experiments and an application to ACTG175 have been conducted to compare our proposed estimator with the others and the results indicates a better performance of our proposed method.


\bibliographystyle{unsrt}
\bibliography{ms}




\end{document}